\documentclass[11pt]{article}
\usepackage{amssymb,amsthm}

\title{Operator-valued Riemann--Hilbert problem for \\
  correlation functions of the $XXZ$ spin chain}
\author{Yasuhiro Fujii\footnotemark[2]\ \ and Miki Wadati}
\date{\textit{\small Department of Physics, Graduate School of Science,
  University of Tokyo,\\  Hongo 7--3--1, Bunkyo-ku, Tokyo 113--0033, Japan}}

\newcommand{\rmi}{\mbox{i}}
\newcommand{\rme}{\mbox{e}}
\newcommand{\rmd}{\mbox{d}}
\newcommand{\rmii}{\mbox{\scriptsize i}}
\newcommand{\der}{\partial}
\renewcommand{\epsilon}{\varepsilon}
\renewcommand{\phi}{\varphi}

\theoremstyle{plain}
\newtheorem*{theorem}{Theorem}
\theoremstyle{definition}
\newtheorem{lemma}{Lemma}
\newtheorem{definition}{Definition}

\newcommand{\vac}{\mbox{vac}}
\newcommand{\2}{\mbox{$\frac{1}{2\pi}$}}
\newcommand{\va}{\mbox{\boldmath$a$}}
\newcommand{\vb}{\mbox{\boldmath$b$}}
\newcommand{\vA}{\mbox{\boldmath$A$}}
\newcommand{\vB}{\mbox{\boldmath$B$}}
\newcommand{\val}{\mbox{\boldmath$\alpha$}}
\newcommand{\vbe}{\mbox{\boldmath$\beta$}}
\newcommand{\vx}{\mbox{\boldmath$\chi$}}
\newcommand{\inn}{\mbox{\scriptsize int}}
\newcommand{\ext}{\mbox{\scriptsize ext}}
\newcommand{\vL}{\mbox{\boldmath$L$}}
\newcommand{\vl}{\mbox{\boldmath$l$}}
\newcommand{\vY}{\mbox{\boldmath$\Psi$}}
\newcommand{\vX}{\mbox{\boldmath$X$}}
\newcommand{\tr}{\mbox{tr}}

\begin{document}

\maketitle
\sloppy
\footnotetext[2]{e-mail: \texttt{fujii@monet.phys.s.u-tokyo.ac.jp}}

\begin{abstract}
  The generating functional of correlation functions
  for the $XXZ$ spin chain is considered in the thermodynamic limit.
  We derive a system of integro-difference equations
  that prescribe this functional.
  On the basis of this system
  we establish the operator-valued Riemann--Hilbert problem
  for correlation functions of the $XXZ$ spin chain.
\end{abstract}

\section{Introduction}

It has been shown by Korepin \textit{et al}
that many important problems arising from analysis of correlation functions
of quantum solvable models
are reduced to classical inverse scattering problems \cite{KBI}.
In particular, for the $XXX$ spin chain,
the generating functional of correlation functions is represented by
the Fredholm determinants \cite{KIEU},
and a special correlation function,
the so-called \textit{Ferromagnetic-String-Formation-Probability}
(FSFP), is proved to be connected with
an operator-valued Riemann--Hilbert problem \cite{FItK}.
With the help of the solutions of this problem
one can compute the large distance asymptotic form of the FSFP \cite{Pr}.
Physically, this makes clear
the probability of finding a ferromagnetic string
of adjacent parallel spins for a given value of the magnetic field
in the $XXX$ spin chain.

In this paper,
we consider the spin-$\frac{1}{2}$ $XXZ$ Heisenberg chain
in the thermodynamic limit.
The Hamiltonian is defined by
\begin{equation}
  H_{XXZ} = \sum_{n\in\mathbb{Z}}
  (\sigma_n^x\sigma_{n+1}^x+\sigma_n^y\sigma_{n+1}^y
  +\cos 2\eta\sigma_n^z\sigma_{n+1}^z-h\sigma_n^z)
\end{equation}
where $\sigma_n^x$, $\sigma_n^y$ and $\sigma_n^z$
are the Pauli matrices acting on the $n$-th site
and $h$ is an external magnetic field.
The anisotropy $\cos 2\eta$ implies the critical regime.
Any correlation function of the model can be obtained
by means of the generating functional $Q^{(m)}(\alpha)$ \cite{EFIK}.
For example, a two-point correlation function
$\langle\sigma_m^z\sigma_1^z\rangle$ is given by
\begin{equation}
  \langle\sigma_m^z\sigma_1^z\rangle =
  2\Delta_m\left.\frac{\der^2 Q^{(m)}(\alpha)}{\der^2\alpha}\right|_{\alpha=0}
  -4\left.\frac{\der Q^{(1)}(\alpha)}{\der\alpha}\right|_{\alpha=0}-1
\end{equation}
with the lattice Laplacian defined by
$\Delta_m f(m)=f(m)-2f(m-1)+f(m-2)$.
This generating functional is
represented by the Fredholm determinants:
\begin{equation}
  \label{Q}
  Q^{(m)}(\alpha) =
  \frac{\langle\vac|\det(1-\2 V^{(m)})|\vac\rangle}
  {\langle\vac|\det(1-\2 V^{(m)}|_{\alpha=0})|\vac\rangle}.
\end{equation}
Here the kernel is of the form:
\begin{eqnarray}
  \label{preV}
  V^{(m)}(\lambda,\mu) &=&
  \frac{\sin 2\eta}{\sinh(\lambda-\mu)}
  \left(\frac{\exp(\alpha-\phi_3(\lambda)+\phi_4(\mu))
      e^{-1}(\lambda)e_1(\mu)+1}{\sinh(\lambda-\mu+2\rmi\eta)}
  \right. \nonumber \\
  && \qquad
  -\left.\frac{e_2^{-1}(\lambda)e_2(\mu)
      +\exp(\alpha-\phi_3(\lambda)+\phi_4(\mu))}
    {\sinh(\mu-\lambda+2\rmi\eta)}\right).
\end{eqnarray}
From now on the index $m$ is omitted if unnecessary.
The integration contour that appears in the Fredholm determinants
lies on the real axis $[-\Lambda,\Lambda]$,
where a boundary value $\Lambda$ is called the Fermi energy
and depends on the anisotropy parameter $\eta$
and the magnetic field $h$.
The operators $e_1(\lambda)$ and $e_2(\lambda)$ are defined by
\begin{equation}
  e_1(\lambda) =
  \left(\frac{\sinh(\lambda+\rmi\eta)}{\sinh(\lambda-\rmi\eta)}\right)^m
  \rme^{-\phi_1(\lambda)}
  \qquad
  e_2(\lambda) =
  \left(\frac{\sinh(\lambda+\rmi\eta)}{\sinh(\lambda-\rmi\eta)}\right)^m
  \rme^{\phi_2(\lambda)}.
\end{equation}
The operators $\phi_j(\lambda)$ ($j=1,\ldots,4$)
are bosonic quantum fields called the \textit{dual fields}.
They are decomposed into the momentum and coordinate fields:
\begin{equation}
  \phi_j(\lambda) = p_j(\lambda)+q_j(\lambda)
  \qquad
  (j=1,\ldots,4)
\end{equation}
whose commutation relations are
\begin{eqnarray}
  \label{com}
  [p_j(\lambda),p_k(\mu)] &=& [q_j(\lambda),q_k(\mu)] = 0
  \nonumber \\\
  [p_j(\lambda),q_k(\mu)] &=&
  U_{jk}h(\lambda,\mu)+U_{kj}h(\mu,\lambda)
  \qquad
  (j,k=1,\ldots,4)
\end{eqnarray}
with
\begin{equation}
  U = -\left(
    \begin{array}{cccc}
      1 & 0 & 1 & 0 \\
      0 & 1 & 0 & 1 \\
      0 & 1 & 1 & 1 \\
      1 & 0 & 1 & 1
    \end{array}
  \right)
  \qquad
  h(\lambda,\mu) =
  \log\frac{\sinh(\lambda-\mu+2\rmi\eta)}{\rmi\sin 2\eta}.
\end{equation}
Notice that the dual fields are commutative:
$[\phi_j(\lambda),\phi_k(\mu)]=0$ ($j,k=1,\ldots,4$).
They only produce vacuum expectation values
according to the vacuum states defined by
\begin{equation}
  \langle\vac|\vac\rangle = 1
  \qquad
  \langle\vac|q_j(\lambda) =
  p_j(\lambda)|\vac\rangle = 0.
  \qquad
  (j=1,\ldots,4)
\end{equation}

A special correlation function FSFP for the $XXZ$ spin chain
is defined by $P(m)=Q^{(m)}(-\infty)$,
where $m$ gives the length of a ferromagnetic string \cite{EFIK,EFItK1}.
In this case the expression of kernel (\ref{preV}) is reduced to
a simple form and enables us to formulate
the related operator-valued Riemann--Hilbert problem
in the same way as for the $XXX$ spin chain \cite{EFItK1}.
Furthermore,
in the limit of strong magnetic field
$h\rightarrow h_c=4\cos^2\eta$,
the kernel is related with the $\tau$-function
of the Painlev\'{e} $V$ equation \cite{EFItK2}.
On the basis of this fact
we can evaluate any correlation function of the $XXZ$ spin chain
under an external strong magnetic field.
Similar results hold in the asymmetric $XXZ$ spin chain
that is a non-hermitian generalization
of the $XXZ$ spin chain \cite{F1,F2}.

The aim of the paper is to derive a system of integro-difference equations
that prescribe the generating functional $Q^{(m)}(\alpha)$
and to establish
the associated operator-valued Riemann--Hilbert problem.
We \textit{do not} take either
the FSFP limit ($\alpha\rightarrow-\infty$)
or the strong magnetic field limit ($h\rightarrow h_c$),
and treat directly the kernel (\ref{preV})
that contains the dual fields with four-species.
This kernel is not of the desired type, however,
it can be transformed into an integral operator
that satisfies the integrable condition (see (\ref{ab0})).
Thus the computation of the generating functional
can be reduced to the solution of
operator-valued Riemann--Hilbert problems.
In section~\ref{sec:io}
we rewrite the kernel (\ref{preV})
and introduce some integral operators.
In particular the resolvent is expressed by
products of two vectors (see Lemma~\ref{lem:R}).
In section~\ref{sec:de}
a system of integro-difference equations is derived.
On the basis of this system
we establish the operator-valued Riemann--Hilbert problem
associated with the generating functional (\ref{Q})
in section~\ref{sec:rhp}.
Section~\ref{sec:con}, the last section,
is devoted to concluding remarks.

\setcounter{equation}{0}
\section{Integral operators}
\label{sec:io}

In this section we transform the kernel (\ref{preV})
into an integral operator that satisfies
the integrable condition (\ref{ab0})
and introduce some useful integral operators.
\begin{definition}
  We rewrite the kernel for the Fredholm determinant representation
  of the generating functional (\ref{Q}) as 
  \begin{equation}
    \label{V}
    V(x,y) =
    \rmi\int_0^\infty\frac{\rmd s}{x-y}
    \sum_{j=1}^4 a_j(x|s)b_j(y|s)
  \end{equation}
  where vectors $\va(x|s)$ and $\vb(x|s)$ are defined by
  \begin{eqnarray}
    \va(x|s) &=&
    \left(\rmi(q-q^{-1})\frac{qx-1}{x-q}\right)^{1/2}
    \left(
      \begin{array}{c}
        x^{m/2}\exp(-\rmi q\frac{qx-1}{x-q}s) \\
        x^{-m/2}\exp(-\rmi q\frac{qx-1}{x-q}s
        +\alpha+\phi_1(x)-\phi_3(x)) \\
        x^{-m/2}\exp(\rmi q^{-1}\frac{qx-1}{x-q}s-\phi_2(x)) \\
        x^{m/2}\exp(\rmi q^{-1}\frac{qx-1}{x-q}s
        +\alpha-\phi_3(x))
      \end{array}
    \right)
    \nonumber \\
    \\
    \vb^t(x|s) &=&
    \left(\rmi(q-q^{-1})\frac{qx-1}{x-q}\right)^{1/2}
    \left(
      \begin{array}{c}
        -x^{-m/2}\exp(\rmi q^{-1}\frac{qx-1}{x-q}s) \\
        -x^{m/2}\exp(\rmi q^{-1}\frac{qx-1}{x-q}s
        -\phi_1(x)+\phi_4(x)) \\
        x^{m/2}\exp(-\rmi q\frac{qx-1}{x-q}s+\phi_2(x)) \\
        x^{-m/2}\exp(-\rmi q\frac{qx-1}{x-q}s+\phi_4(x))
      \end{array}
    \right).
  \end{eqnarray}
  The superscript $t$ indicates the transposition of a vector.
  The integration contour for variables $x,y$ runs
  on the unit circle anti-clockwise: $C:x=\rme^{\rmii\theta}$
  $(-\psi<\theta<2\pi+\psi)$.
  The endpoints are denoted by $\zeta_\pm=\rme^{\pm\rmii\psi}$.
  The dual fields still obey the commutation relations
  (\ref{com}) with
  \begin{equation}
    h(x,y) =
    \log\frac{q\sqrt{\frac{qx-1}{x-q}\frac{y-q}{qy-1}}
      -q^{-1}\sqrt{\frac{x-q}{qx-1}\frac{qy-1}{y-q}}}{q-q^{-1}}.
  \end{equation}
\end{definition}
The original kernel (\ref{preV}) is obtained
by using transformations
\begin{equation}
  x = \frac{\sinh(\lambda+\rmi\eta)}{\sinh(\lambda-\rmi\eta)}
  \qquad
  q = \rme^{2\rmii\eta}
  \qquad
  \psi = \rmi\log\frac{\sinh(\Lambda-\rmi\eta)}{\sinh(\Lambda+\rmi\eta)}
\end{equation}
and an identity that is valid for $\frac{\pi}{2}<\eta<\pi$:
\begin{equation}
  \int_0^\infty\rmd s \exp
  \left(-\rmi q\frac{qx-1}{x-q}s+\rmi q^{-1}\frac{qy-1}{y-q}s\right)
  = \frac{-\rmi}{q\frac{qx-1}{x-q}-q^{-1}\frac{qy-1}{y-q}}.
\end{equation}
Here an extra factor that remains the Fredholm determinants unchanged
appear, however, it can be ignored
because it has no influence on the representation
of the generating functional (\ref{Q}).
We remark that the vectors $\va(x|s)$ and $\vb(x|s)$ satisfy
\begin{equation}
  \label{ab0}
  \int_0^\infty\rmd s \sum_{j=1}^4 a_j(x|s)b_j(x|s) = 0.
\end{equation}
This relation is called the \textit{integrable condition}
and assures that the kernel (\ref{V}) is well-defined
for any $x,y\in\mathbb{C}$.
By virtue of this condition
we can formulate the operator-valued Riemann--Hilbert problem
in section~\ref{sec:rhp}.

Hereafter some matrices that contain integral operators will appear.
We call them  the \textit{matrix integral operators} 
and denote them by bold letters,
except that ``$1$'' indicates the product of
the delta-function and the unit matrix.
The following convention for matrix integral operators
is adopted unless it invites confusion:
\begin{equation}
  (\mbox{\boldmath$E$}\mbox{\boldmath$F$})_{jk}(x,y|s,t) =
  \int_C \rmd z \int_0^\infty \rmd r
  \sum_n E_{jn}(x,z|s,r)F_{nk}(z,y|r,t).
\end{equation}
\begin{definition}
  We introduce the vectors $\val(x|s)$ and $\vbe(x|s)$ via
  \begin{equation}
    \label{albe}
    ((1-\2 V)\val)(x|s) = \va(x|s)
    \qquad
    (\vbe(1-\2 V))(x|s) = \vb(x|s)
  \end{equation}
  and define the \textit{resolvent} $R(x,y)$ through
  \begin{equation}
    \label{vr}
    (1-\2 V)(1+\2 R) = (1+\2 R)(1-\2 V) = 1.
  \end{equation}
\end{definition}
Note that one of these relations can be derived
by letting $(1+\2 R)$ act on both sides of the other relation
from the left or the right.
Thus $(1-\2 V)$ and $(1+\2 R)$ are commutative.
\begin{lemma}
  \label{lem:R}
  The resolvent is represented by
  \begin{equation}
    \label{R}
    R(x,y) = \rmi\int_0^\infty \frac{\rmd s}{x-y}
    \sum_{j=1}^4 \alpha_j(x|s)\beta_j(y|s).
  \end{equation}
\end{lemma}
\begin{proof}
  The defining relation of the resolvent is rewritten as
  $(1-\2 V)R=V$.
  By using the identity $(x-y)=(x-z)+(z-y)$
  this relation is deformed into
  \begin{eqnarray}
    (x-y)R(x,y) &-&\frac{\rmi}{2\pi}\int_C\rmd z \int_0^\infty\rmd s
    \sum_{j=1}^4 a_j(x|s)b_j(z|s)R(z,y)
    \nonumber \\
    &-&\frac{1}{2\pi}\int_C \rmd z V(x,z)(z-y)R(z,y)
    \nonumber \\ &=&
    \rmi\int_0^\infty\rmd s \sum_{j=1}^4 a_j(x|s)b_j(y|s).
  \end{eqnarray}
  Move the second term in the left side to the right side
  and recall the defining relations (\ref{albe})--(\ref{vr}).
  The representation (\ref{R}) is thus obtained.
\end{proof}
\begin{definition}
  \label{defab}
  We define the $4\times 4$ matrix integral operators
  with variables $s,t$ by
  \begin{eqnarray}
    A_{jk}(s,t) &=& \rmi\int_C\frac{\rmd x}{x}\alpha_j(x|s)b_k(x|t)
    \\
    B_{jk}(s,t) &=& \rmi\int_C\frac{\rmd x}{x}a_j(x|s)\beta_k(x|t).
    \qquad
    (j,k=1,\ldots,4)
  \end{eqnarray}
\end{definition}
\begin{lemma}
  \label{lem:ab}
  The integral operator $\vB(s,t)$ is the resolvent
  of $\vA(s,t)$. Namely,
  \begin{equation}
    (1-\2\vA)(1+\2\vB) = (1+\2\vB)(1-\2\vA) = 1.
  \end{equation}
\end{lemma}
\begin{proof}
  The action of $(1-\2 V)$ on $\frac{1}{x}\val(x|s)$
  is computed as
  \begin{eqnarray}
    \label{aoverx}
    \lefteqn{\frac{1}{x}\val(x|s)
      -\frac{1}{2\pi}\int_C\frac{\rmd y}{y}V(x,y)\val(y|s)}
    \nonumber \\ &=&
    \frac{1}{x}\va(x|s)
    +\frac{1}{2\pi}\int_C\rmd y\left(\frac{1}{x}-\frac{1}{y}\right)
    V(x,y)\val(y|s)
    \nonumber \\ &=&
    \frac{1}{x}((1-\2\vA)\va)(x|s).
  \end{eqnarray}
  By virtue of this relation we obtain
  \begin{eqnarray}
    A_{jk}(s,t) &=&
    \rmi\int_C\frac{\rmd x}{x}\alpha_j(x|s)(\beta_k(1-\2 V))(x|t)
    \nonumber \\ &=&
    \rmi\int_C\rmd x \beta_k(x|t)
    \left(\frac{1}{x}\alpha_j(x|s)
      -\frac{1}{2\pi}\int_C\frac{\rmd y}{y}V(x,y)\alpha_j(y|s)\right)
    \nonumber \\ &=&
    \rmi\int_C\frac{\rmd x}{x}\beta_k(x|t)
    ((1-\2\vA)\va)_j(x|s)
    \nonumber \\ &=&
    ((1-\2\vA)\vB)_{jk}(s,t).
    \qquad
    (j,k=1,\ldots,4)
  \end{eqnarray}
  This means $(1-\2\vA)\vB=\vA$
  and therefore implies the lemma.
  The proof has been completed.
\end{proof}
Applying Lemma~\ref{lem:ab} to (\ref{aoverx})
we can obtain another representation of $\frac{1}{x}\va(x|s)$:
\begin{equation}
  \label{xa}
  \frac{1}{x}\va(x|s) =
  \frac{1}{x}((1+\2\vB)\val)(x|s)
  -\frac{1}{2\pi}\int_C\frac{\rmd y}{y}V(x,y)((1+\2\vB)\val)(y|s).
\end{equation}
We remark that the kernel $V(x,y)$
commutes with $\vA(s,t)$ and $\vB(s,t)$
because variables are different.
This relation plays an important role in the next section.

\setcounter{equation}{0}
\section{Integro-difference equations}
\label{sec:de}

In order to make clear the properties of the unknown vectors
$\val^{(m)}(x|s)$ and $\vbe^{(m)}(x|s)$
we compute their dependences on $m$ and $\psi$,
and derive a system of integro-difference equations for them.
This helps us to formulate an operator-valued
Riemann-Hilbert problem for the $XXZ$ spin chain.
\begin{lemma}
  \label{lem:dm}
  The vectors $\val^{(m)}(x|s)$ and $\vbe^{(m)}(x|s)$
  obey the following integro-difference equations with respect to $m$:
  \begin{eqnarray}
    \label{dval}
    \val^{(m+1)}(x|s) &=&
    ((\sqrt{x}\gamma_1+\frac{1}{\sqrt{x}}
    (1-\2\vA^{(m+1)})\gamma_2(1+\2\vB^{(m)}))
    \val^{(m)})(x|s)
    \\
    \label{dvbe}
    \vbe^{(m+1)}(x|s) &=&
    (\vbe^{(m)}(\sqrt{x}\gamma_2+\frac{1}{\sqrt{x}}
    (1-\2\vA^{(m)})\gamma_1(1+\2\vB^{(m+1)})))(x|s)
  \end{eqnarray}
  where $\gamma_1$ and $\gamma_2$ are $4\times 4$ diagonal matrices:
  \begin{equation}
    \gamma_1 = \frac{1}{2}(1+\sigma^z\otimes\sigma^z)
    \qquad
    \gamma_2 = \frac{1}{2}(1-\sigma^z\otimes\sigma^z).
  \end{equation}
\end{lemma}
\begin{proof}
  By definitions of $\va^{(m)}(x|s)$ and $\vb^{(m)}(x|s)$
  their dependences on $m$ are
  \begin{eqnarray}
    \va^{(m+1)}(x|s) &=&
    (\sqrt{x}\gamma_1+\frac{1}{\sqrt{x}}\gamma_2)\va^{(m)}(x|s)
    \\
    \vb^{(m+1)}(x|s) &=&
    \vb^{(m)}(x|s)(\sqrt{x}\gamma_2+\frac{1}{\sqrt{x}}\gamma_1).
  \end{eqnarray}
  Similarly the kernel $V^{(m+1)}(x,y)$ is connected
  with $V^{(m)}(x,y)$ as
  \begin{equation}
    \label{reV}
    V^{(m+1)}(x,y) =
    \sqrt{\frac{x}{y}}\left(
      V^{(m)}(x,y)-\rmi\int_0^\infty\frac{\rmd s}{x}
      \vb^{(m)}(y|s)\gamma_2\va^{(m)}(x|s)\right).
  \end{equation}
  Using these recursion relations we obtain
  \begin{eqnarray}
    \lefteqn{\val^{(m+1)}(x|s)-\va^{(m+1)}(x|s)}
    \nonumber \\ &=&
    \frac{1}{2\pi}(V^{(m+1)}\val^{(m+1)})(x|s)
    \nonumber \\ &=&
    \frac{\sqrt{x}}{2\pi}\int_C\frac{\rmd y}{\sqrt{y}}
    V^{(m)}(x,y)\val^{(m+1)}(y|s)
    -\frac{1}{2\pi\sqrt{x}}(\vA^{(m+1)}\gamma_2\va^{(m)})(x|s).
  \end{eqnarray}
  It thus follows that
  \begin{eqnarray}
    \lefteqn{\frac{1}{\sqrt{x}}\val^{(m+1)}(x|s)
      -\frac{1}{2\pi}\int_C\frac{\rmd y}{\sqrt{y}}
      V^{(m)}(x,y)\val^{(m+1)}(y|s)}
    \nonumber \\ &=&
    \frac{1}{\sqrt{x}}\va^{(m+1)}(x|s)
    -\frac{1}{2\pi x}(\vA^{(m+1)}\gamma_2\va^{(m)})(x|s)
    \nonumber \\ &=&
    \gamma_1\va^{(m)}(x|s)
    +\frac{1}{x}((1-\2\vA^{(m+1)})\gamma_2\va^{(m)})(x|s)
    \nonumber \\ &=&
    ((1-\2 V^{(m)})\gamma_1\val^{(m)})(x|s)
    +\frac{1}{x}((1-\2\vA^{(m+1)})\gamma_2(1+\2\vB^{(m)})\val^{(m)})(x|s)
    \nonumber \\ &&
    -\frac{1}{2\pi}\int_C\frac{\rmd y}{y}V^{(m)}(x,y)
    ((1-\2\vA^{(m+1)})\gamma_2(1+\2\vB^{(m)})\val^{(m)})(y|s).
  \end{eqnarray}
  In the last equality the relation (\ref{xa}) is used.
  Thus the integro-difference equation (\ref{dval}) is obtained
  by removing the action of $(1-\2 V^{(m)})$.
  In the same way (\ref{dvbe}) can be derived.
\end{proof}
\begin{lemma}
  \label{lem:dy}
  The derivative of the vectors $\val(x|s)$ and $\vbe(x|s)$
  with respect to $\psi$ are
  \begin{eqnarray}
    \label{dera}
    \der_\psi\val(x|s) &=&
    -\frac{1}{2\pi}\sum_{\epsilon=\pm}\left(
      \frac{\zeta_\epsilon}{x-\zeta_\epsilon}\val(\zeta_\epsilon|s)
      \int_0^\infty\rmd t
      \sum_{j=1}^4 \alpha_j(x|t)\beta_j(\zeta_\epsilon|t)
    \right)
    \\
    \label{derb}
    \der_\psi\vbe(x|s) &=&
    -\frac{1}{2\pi}\sum_{\epsilon=\pm}\left(
      \frac{\zeta_\epsilon}{\zeta_\epsilon-x}\vbe(\zeta_\epsilon|s)
      \int_0^\infty\rmd t
      \sum_{j=1}^4 \alpha_j(\zeta_\epsilon|t)\beta_j(x|t)
    \right).
  \end{eqnarray}
\end{lemma}
\begin{proof}
  We show the differential equation (\ref{dera}).
  The other equation (\ref{derb}) can be derived in the same way.
  According to the identity
  \begin{equation}
    -\rmi\der_\psi\int_C\rmd x f(x) =
    \sum_{\epsilon=\pm}\zeta_\epsilon f(\zeta_\epsilon)
  \end{equation}
  the differentiation of (\ref{albe}) is computed as
  \begin{equation}
    \der_\psi\val(x|s) -\frac{\rmi}{2\pi}\sum_{\epsilon=\pm}
    \zeta_\epsilon V(x,\zeta_\epsilon)\val(\zeta_\epsilon|s)
    -\frac{1}{2\pi}(V\der_\psi\val)(x|s) = 0.
  \end{equation}
  By using the resolvent this is reduced to
  \begin{equation}
   \der_\psi\val(x|s) =
   \frac{\rmi}{2\pi}\sum_{\epsilon=\pm}
   \zeta_\epsilon R(x,\zeta_\epsilon)\val(\zeta_\epsilon|s).
  \end{equation}
  Because of Lemma~\ref{lem:R} we obtain (\ref{dera}).
\end{proof}

The following lemma is not directly related with
the formulation of operator-valued Riemann--Hilbert problems
but useful for evaluating the large $m$ asymptotic form
of the generating functional $Q^{(m)}(\alpha)$:
\begin{lemma}
  \label{lem:V}
  The Fredholm determinant $\det(1-\2 V^{(m)})$
  satisfies the following recursion relation with respect to $m$:
  \begin{equation}
    \frac{\det(1-\2 V^{(m+1)})}{\det(1-\2 V^{(m)})} =
    \exp\tr\log(1+\2\gamma_2\vB^{(m)})
  \end{equation}
  where the trace for matrix integral operators is defined by
  \begin{equation}
    \tr(\mbox{\boldmath$K$}^n) =
    \int_0^\infty \prod_{k=1}^n\rmd s_k\sum_{j_1,\ldots,j_n=1}^4
    K_{j_1,j_2}(s_1,s_2)\cdots K_{j_n,j_1}(s_n,s_1).
  \end{equation}
\end{lemma}
\begin{proof}
  Due to the relation (\ref{reV}) it follows that
  \begin{eqnarray}
    \lefteqn{\delta(x-y)
      -\frac{1}{2\pi}\sqrt{\frac{y}{x}}V^{(m+1)}(x,y)}
    \nonumber \\ &=&
    \delta(x-y)-\frac{1}{2\pi}\left(
      V^{(m)}(x,y)-\rmi\int_0^\infty\frac{\rmd s}{x}
      (\vbe^{(m)}(1-\2 V^{(m)}))(y|s)\gamma_2\va^{(m)}(x|s)\right)
    \nonumber \\ &=&
    ((1+\2 G^{(m)})(1-\2 V^{(m)}))(x,y)
  \end{eqnarray}
  where
  \begin{equation}
    G^{(m)}(x,y) =
    \rmi\int_0^\infty\frac{\rmd s}{x}
    \vbe^{(m)}(y|s)\gamma_2\va^{(m)}(x|s).
  \end{equation}
  Based on this recursion relation
  the Fredholm determinants of both sides are given by
  \begin{equation}
    \label{vgv}
    \det(1-\2 V^{(m+1)}) =
    \det(1+\2 G^{(m)})\det(1-\2 V^{(m)}).
  \end{equation}
  Here we note that $\sqrt{y/x}$ in the left side
  has no influence on the Fredholm determinant.
  Take the logarithm of $\det(1+\2 G)$:
  \begin{equation}
    \label{dettr}
    \log\det(1+\2 G) = \tr\log(1+\2 G) =
    \sum_{n=1}^\infty\frac{(-1)^{n-1}}{n(2\pi)^n}\tr(G^n).
  \end{equation}
  The trace of $G^n$ can be expressed in terms of $\vB$:
  \begin{eqnarray}
    \label{trg}
    \tr(G^n) &=&
    \rmi^n \int_C \prod_{k=1}^n\frac{\rmd x_k}{x_k}
    \int_0^\infty \prod_{k=1}^n \rmd s_k
    \nonumber \\ &&
    \times\sum_{j_1,\ldots,j_n=2,3}
    a_{j_1}(x_1|s_1)\beta_{j_1}(x_2|s_1)
    a_{j_2}(x_2|s_2)\beta_{j_2}(x_3|s_2)
    \cdots a_{j_n}(x_n|s_n)\beta_{j_n}(x_1|s_n)
    \nonumber \\ &=&
    \int_0^\infty \prod_{k=1}^n \rmd s_k
    \sum_{j_1,\ldots,j_n=2,3}
    B_{j_1,j_n}(s_1,s_n)B_{j_2,j_1}(s_2,s_1)
    \cdots B_{j_n,j_{n-1}}(s_n,s_{n-1})
    \nonumber \\ &=&
    \tr((\gamma_2\vB)^n).
  \end{eqnarray}
  Equation (\ref{vgv}) with (\ref{dettr}) and (\ref{trg})
  proves the lemma.
\end{proof}

\setcounter{equation}{0}
\section{Operator-valued Riemann--Hilbert problem}
\label{sec:rhp}

In this section we establish
the operator-valued Riemann--Hilbert problem
associated with the generating functional (\ref{Q}).
Let us consider the $4\times 4$ matrix integral operator
$\vx^{(m)}(z|s,t)$ that has the following properties.
Hereafter not only the index $m$ but also
variables $s$, $t$ are omitted if unnecessary.
\begin{enumerate}
\item $\vx(z)$ is analytic for $z\in\mathbb{C}\backslash C$.
\item Let $\vx_{\ext}(z)$ and $\vx_{\inn}(z)$ be the limits of $\vx(z)$
  from outside and inside of the unit circle, respectively.
  It then follows that
  \begin{equation}
    \label{extint}
    \vx_{\ext}(z) = \vx_{\inn}(z)\vL(z)
    \qquad
    (z\in C)
  \end{equation}
  where the \textit{conjugation matrix} $\vL(z)$ is given by
  \begin{equation}
    \vL(z) = 1-\vl(z)
    \qquad
    l_{jk}(z|s,t) = a_j(z|s)b_k(z|t).
    \qquad
    (j,k=1,\ldots,4)
  \end{equation}
\item $\vx(\infty)=1$.
\end{enumerate}
We point out that the conjugation matrix $\vL(z)$
is the $4\times 4$ matrix integral operator with variables $s,t$.
For example, the condition (\ref{extint}) means
\begin{equation}
  (\vx_{\ext})_{jk}(z|s,t) =
  \int_0^\infty\rmd r\sum_{n=1}^4
  (\vx_{\inn})_{jn}(z|s,r)L_{nk}(z|r,t).
  \qquad
  (j,k=1,\ldots,4)
\end{equation}

The connection of the operator-valued Riemann-Hilbert problem 1--3
to Lemmas~\ref{lem:R}--\ref{lem:V} in the previous section
is summarized in the following theorem:
\begin{theorem}
  Suppose that the solution of the Riemann--Hilbert problem 1--3
  exists and is unique.
  Define new vectors by
  \begin{equation}
    \val(z|s) = \int_0^\infty\rmd t\vx(z|s,t)\va(z|t)
    \qquad
    \vbe(z|s) = \int_0^\infty\rmd t\vb(z|t)\vx^{-1}(z|t,s).
  \end{equation}
  and introduce the integral operators $\vA(s,t)$ and $\vB(s,t)$
  as Definition~\ref{defab} again.
  Then they satisfy Lemmas~\ref{lem:ab}--\ref{lem:dy}.
\end{theorem}
\begin{proof}
  We start from the proof of Lemma~\ref{lem:ab}.
  Using the definition of $\vA(s,t)$
  and the \textit{canonical integral representation} of $\vx(z)$
  (see \cite{KBI,FItK,EFItK1})
  \begin{equation}
    \vx(z) =
    1+\frac{1}{2\pi\rmi}\int_C\frac{\rmd\zeta}{\zeta-z}
    \vx_{\inn}(\zeta)\vl(\zeta)
  \end{equation}
  we obtain
  \begin{equation}
    \vx(0) =
    1-\frac{\rmi}{2\pi}\int_C\frac{\rmd z}{z}\vx_{\inn}(z)\vl(z) =
    1-\frac{1}{2\pi}\vA.
  \end{equation}
  Since $\vl^2(z)=0$
  (which is clear from the integrable condition (\ref{ab0}))
  there exists the inverse of the conjugation matrix:
  $\vL^{-1}(z)=1+\vl(z)$.
  The canonical integral representation of $\vx^{-1}(z)$
  is thus given by
  \begin{equation}
    \label{xb}
    \vx^{-1}(0) =
    1+\frac{\rmi}{2\pi}\int_C\frac{\rmd z}{z}
    \vl(z)\vx_{\inn}^{-1}(z) =
    1+\frac{1}{2\pi}\vB.
  \end{equation}
  The identity $\vx(0)\vx^{-1}(0)=\vx^{-1}(0)\vx(0)=1$
  corresponds to Lemma~\ref{lem:ab}.

  Let us compute the integro-difference relation of $\vx^{(m)}(z)$
  with respect to $m$ and prove Lemma~\ref{lem:dm}.
  We introduce the following integral operator
  \begin{equation}
    \vY^{(m)}(z) = \vx^{(m)}(z)(\gamma_1+z^{-m}\gamma_2).
  \end{equation}
  This obeys
  \begin{equation}
    \vY_{\ext}^{(m)}(z) = \vY_{\inn}^{(m)}(z)\vL_0(z)
    \qquad
    (z\in C)
  \end{equation}
  with the conjugation matrix
  \begin{equation}
    \vL_0(z) =
    (\gamma_1+z^m\gamma_2)\vL^{(m)}(z)(\gamma_1+z^{-m}\gamma_2).
  \end{equation}
  From the definitions of $\va^{(m)}(z)$, $\vb^{(m)}(z)$ and $\vL^{(m)}(z)$,
  we see that $\vL_0(z)$ is independent of $m$.
  Hence applying Liouville's theorem we have
  \begin{equation}
    \vY^{(m+1)}(z)(\vY^{(m)}(z))^{-1} =
    \gamma_1+\frac{1}{z}\vx^{(m+1)}(0)\gamma_2(\vx^{(m)}(0))^{-1}
  \end{equation}
  which implies
  \begin{equation}
    \vx^{(m+1)}(z)(\sqrt{z}\gamma_1+\frac{1}{\sqrt{x}}\gamma_2) =
    (\sqrt{z}\gamma_1+\frac{1}{\sqrt{x}}\vx^{(m+1)}(0)\gamma_2
    (\vx^{(m)}(0))^{-1})\vx^{(m)}(z).
  \end{equation}
  By letting $\va^{(m)}(z|s)$ act on this relation from the right
  the integro-difference equation (\ref{dval}) is derived.
  Similarly (\ref{dvbe}) is obtained.
  Thus Lemma~\ref{lem:dm} has been proved.

  We show Lemma~\ref{lem:dy}.
  In the same way as the FSFP case \cite{EFItK1}
  and the impenetrable Bose gas case \cite{JMMS},
  in the neighborhood of $C$,
  the integral operator $\vx(z)$ can be decomposed into
  \begin{equation}
    \vx(z) = \widehat{\vx}(z)\vx_0(z)
  \end{equation}
  where $\widehat{\vx}(z)$ is a single-valued, invertible and analytic
  in the neighborhood of $C$.
  $\vx_0(z)$ is represented by
  \begin{equation}
    \vx_0(z) =
    1+\frac{\rmi}{2\pi}\log\frac{z-\zeta_-}{z-\zeta_+}\vl(z).
  \end{equation}
  Since $\vl^2(z)=0$ its logarithm-derivative is computed as
  \begin{equation}
    \der_\psi\vx_0(z)\vx_0^{-1}(z) =
    -\frac{1}{2\pi}\sum_{\epsilon=\pm}
    \frac{\zeta_\epsilon}{z-\zeta_\epsilon}\vl(z).
  \end{equation}
  Due to this relation and Liouville's theorem
  the logarithm-derivative of $\vx(z)$ is written as
  \begin{equation}
    \label{derx}
    \der_\psi\vx(z)\vx^{-1}(z) =
    \sum_{\epsilon=\pm}\frac{1}{z-\zeta_\epsilon}\vX_\epsilon
  \end{equation}
  with the coefficient $\vX_{\pm}$:
  \begin{eqnarray}
    \vX_\pm &=&
    \lim_{z\rightarrow\zeta_\pm}(z-\zeta_\pm)
    \widehat{\vx}(z)\der_\psi\vx_0(z)\vx_0^{-1}(z)\widehat{\vx}^{-1}(z)
    \nonumber \\ &=&
    -\frac{\zeta_\pm}{2\pi}
    \widehat{\vx}(\zeta_\pm)\vl(\zeta_\pm)
    \widehat{\vx}^{-1}(\zeta_\pm)
    \nonumber \\ &=&
    -\frac{\zeta_\pm}{2\pi}
    \vx(\zeta_\pm)\vl(\zeta_\pm)\vx^{-1}(\zeta_\pm).
  \end{eqnarray}
  By definitions of $\val(z|s)$ and $\vbe(z|s)$
  the elements of $\vX_{\pm}$ are expressed by
  \begin{equation}
    (\vX_\pm)_{jk}(s,t) =
    -\frac{\zeta_\pm}{2\pi}
    \alpha_j(\zeta_\pm|s)\beta_k(\zeta_\pm|t).
    \qquad
    (j,k=1,\ldots,4)
  \end{equation}
  Let $\vx(z)$ act on (\ref{derx}) from the right
  and use this representation of $\vX_{\pm}$.
  The differential equation (\ref{dera}) is thus obtained.
  Similarly (\ref{derb}) can be derived.
  The proof of the theorem has been completed.
\end{proof}

In consequence of Riemann--Hilbert problem 1--3
Lemmas~\ref{lem:R},~\ref{lem:V} certainly hold.
On the basis of these lemmas
we can evaluate the large $m$ asymptotic behaviors
of correlation functions for the $XXZ$ spin chain.

\setcounter{equation}{0}
\section{Concluding remarks}
\label{sec:con}

In the present paper we have derived a system of integro-difference equations
that prescribe the generating functional of correlation functions
and have established an operator-valued
Riemann--Hilbert problem for the $XXZ$ spin chain.
It can be easily checked that,
in the limit $\alpha\rightarrow-\infty$,
our problem reduces to the problem associated with the FSFP 
obtained in \cite{EFItK1}.
We are in a position to evaluate
the long-distance asymptotic behavior
of any correlation function for the $XXZ$ spin chain.
In a forthcoming publication
our Riemann--Hilbert problem will be used
to compute the large $m$ asymptotic form of the generating functional.

In the free-fermionic point $\eta=3\pi/4$
the $XXZ$ spin chain is reduced to the $XXO$ model.
The FSFP of the $XXO$ model is known to be connected with
an Riemann--Hilbert problem
whose conjugation matrix is not an integral operator \cite{DIZ}.
Its large $m$ asymptotic form is computed as follows:
\begin{equation}
  P_{XXO}(m) \sim \left(\frac{h+2}{4}\right)^{m^2/2}
\end{equation}
where $h$ is the magnetic field.
It is interesting to reproduce the same result
from our Riemann--Hilbert problem
by taking the limit $\alpha\rightarrow-\infty$ and $\eta\rightarrow3\pi/4$.
Furthermore,
we can evaluate any correlation function for the $XXO$ model.

Operator-valued Riemann--Hilbert problems appear in several subjects,
apart from the computation of correlation functions
of quantum solvable models.
For example, it is well known that
the classical inverse problem for integrable equations
in $2+1$-dimensions (Davey--Stewartson, KP, etc)
can be expressed as an operator-valued Riemann--Hilbert problem.
The investigation of operator-valued Riemann--Hilbert problems
is thus important and interesting
from standpoints of not only physics but also mathematics.

\section*{Acknowledgment}

The authors thank to Prof. V. E. Korepin and
Dr. T. Tsuchida for useful comments.


\newpage


\begin{thebibliography}{99}
\bibitem{KBI} V. E. Korepin, N. M. Bogoliubov, A. G. Izergin:
  ``\textit{Quantum inverse scattering method and correlation functions}'',
  (Cambridge University Press, 1993, Cambridge).
\bibitem{KIEU} V. E. Korepin, A. G. Izergin, F. H. L. E\ss ler
  and D. B. Uglov:
  \textit{Phys. Lett.} A \textbf{190} (1994) 182--184.
\bibitem{FItK} H. Frahm, A. R. Its and V. E. Korepin:
  \textit{Nucl. Phys.} B \textbf{428} (1994) 694--710.
\bibitem{Pr} H. Frahm, A. R. Its and V. E. Korepin:
  ``\textit{An operator-valued Riemann--Hilbert problem associated
    with the $XXX$ model}'',
  (CRM Proc. Lecture Notes \textbf{9}, AMS, 1996) 133--142.
\bibitem{EFIK} F. H. L. E\ss ler, H. Frahm, A. G. Izergin
  and V. E. Korepin:
  \textit{Commun. Math. Phys.} \textbf{174} (1995) 191--214.
\bibitem{EFItK1} F. H. L. E\ss ler, H. Frahm, A. R. Its
  and V. E. Korepin:
  \textit{Nucl. Phys.} B \textbf{446} (1995) 448--460.
\bibitem{EFItK2} F. H. L. E\ss ler, H. Frahm, A. R. Its
  and V. E. Korepin:
  \textit{J. Phys.} A \textbf{29} (1996) 5619--5626.
\bibitem{F1} Y. Fujii and M. Wadati:
  \textit{J. Phys. Soc. Jpn.} \textbf{68} (1999) 2228--2233.
\bibitem{F2} Y. Fujii and M. Wadati:
  \textit{J. Phys. Soc. Jpn.} \textbf{68} (1999) 3227--3235.
\bibitem{JMMS} M. Jimbo, T. Miwa, Y. Mori and M. Sato:
  \textit{Physica} D \textbf{1} (1980) 80--158.
\bibitem{DIZ} P. Deift, A. R. Its and X. Zhou:
  \textit{Ann. Math.} \textbf{146} (1997) 149--235.
\end{thebibliography}
\end{document}